\documentclass[journal,twoside,web]{ieeecolor}
\usepackage{generic}
\usepackage{amsmath,amssymb,amsfonts}
\usepackage{algorithmic}
\usepackage{graphicx}
\usepackage{textcomp}

\usepackage{amsmath} % assumes amsmath package installed
\usepackage{amssymb}  % assumes amsmath package installed
\usepackage{graphicx}
\usepackage{multirow}
\usepackage{booktabs}
\usepackage{float}
\usepackage{graphicx}
\usepackage{caption}
\usepackage{mathtools}

\usepackage{tabularx}
\usepackage[table]{xcolor}
\usepackage{colortbl}
\usepackage{tablestyles}

\def\BibTeX{{\rm B\kern-.05em{\sc i\kern-.025em b}\kern-.08em
    T\kern-.1667em\lower.7ex\hbox{E}\kern-.125emX}}
\usepackage[style=numeric,sorting=none]{biblatex}
\usepackage{tikz}
\usepackage{textcomp}
\usepackage{hyperref}
\usepackage{lipsum}

\newcommand\copyrighttext{%
  \footnotesize \textcopyright 2022 IEEE. Personal use of this material is permitted.
  Permission from IEEE must be obtained for all other uses, in any current or future
  media, including reprinting/republishing this material for advertising or promotional
  purposes, creating new collective works, for resale or redistribution to servers or
  lists, or reuse of any copyrighted component of this work in other works.
  DOI: \href{<http://tex.stackexchange.com>}{<DOI No.>}}
\newcommand\copyrightnotice{%
\begin{tikzpicture}[remember picture,overlay]
\node[anchor=south,yshift=10pt] at (current page.south) {\fbox{\parbox{\dimexpr\textwidth-\fboxsep-\fboxrule\relax}{\copyrighttext}}};
\end{tikzpicture}%
}

\begin{document}
\title{Gait Events Prediction using Hybrid CNN-RNN-based Deep Learning models through a Single Waist-worn Wearable Sensor}
\author{Muhammad Zeeshan Arshad, Ankhzaya Jamsrandorj, Jinwook Kim, and Kyung-Ryoul Mun% <-this % stops a space
\thanks{This paper revision was submitted for review on January 17, 2022.}
\thanks{M.Z. Arshad and J. Kim are with the Center for Artificial Intelligence, KIST, Seoul, Republic of Korea.}
\thanks{A. Jamsrandorj is with the Department of Human Computer Interface \& Robotics Engineering, University of Science \& Technology, Daejon, Republic of Korea.}
\thanks{K. Mun is with the Center for Artificial Intelligence, KIST, Seoul, Republic of Korea and KHU-KIST Department of Converging Science and Technology, KHU, Seoul, Republic of Korea.}
}
% \thanks{S. B. Author, Jr., was with Rice University, Houston, TX 77005 USA. He is 
% now with the Department of Physics, Colorado State University, Fort Collins, 
% CO 80523 USA (e-mail: author@lamar.colostate.edu).}
% \thanks{T. C. Author is with 
% the Electrical Engineering Department, University of Colorado, Boulder, CO 
% 80309 USA, on leave from the National Research Institute for Metals, 
% Tsukuba, Japan (e-mail: author@nrim.go.jp).}}

\maketitle
\copyrightnotice
\begin{abstract}
Elderly gait is a source of rich information about their physical and mental health condition. As an alternative to the multiple sensors on the lower body parts, a single sensor on the pelvis has a positional advantage and an abundance of information acquirable. This study aimed to explore a way of improving the accuracy of gait event detection in the elderly using a single sensor on the waist and deep learning models. Data was gathered from elderly subjects equipped with three IMU sensors while they walked. The input was taken only from the waist sensor was used to train 16 deep-learning models including CNN, RNN, and CNN-RNN hybrid with or without the Bidirectional and Attention mechanism. The groundtruth was extracted from foot IMU sensors. Fairly high accuracy of 99.73\% and 93.89\% was achieved by the CNN-BiGRU-Att model at the tolerance window of $\pm$6TS ($\pm$6ms) and $\pm$1TS ($\pm$1ms) respectively. Advancing from the previous studies exploring gait event detection, the model showed a great improvement in terms of its prediction error having an MAE of 6.239ms and 5.24ms for HS and TO events respectively at the tolerance window of $\pm$1TS. The results showed that the use of CNN-RNN hybrid models with Attention and Bidirectional mechanisms is promising for accurate gait event detection using a single waist sensor. The study can contribute to reducing the burden of gait detection and increase its applicability in future wearable devices that can be used for remote health monitoring (RHM) or diagnosis based thereon.
\end{abstract}


\section{Introduction}
\label{sec:introduction}
\IEEEPARstart{T}{he} gait of the elderly is abundant with health information on not only their current status but also the potential health risk they are at \cite{studenski2011gait}. On top of physical conditions, even mental conditions like cognitive impairment and dementia can be found in their gait \cite{arshad2021gait, jung2021classifying, verghese2008gait, mielke2013assessing}. Precise measurement of the gait among the elderly allows us to predict and detect their medical crisis at an early stage and to establish an active strategy to prevent unnecessary disease progression.

\begin{figure*}[tb!]
    \centering
    \captionsetup{justification=centering}
    \includegraphics[width=\linewidth]{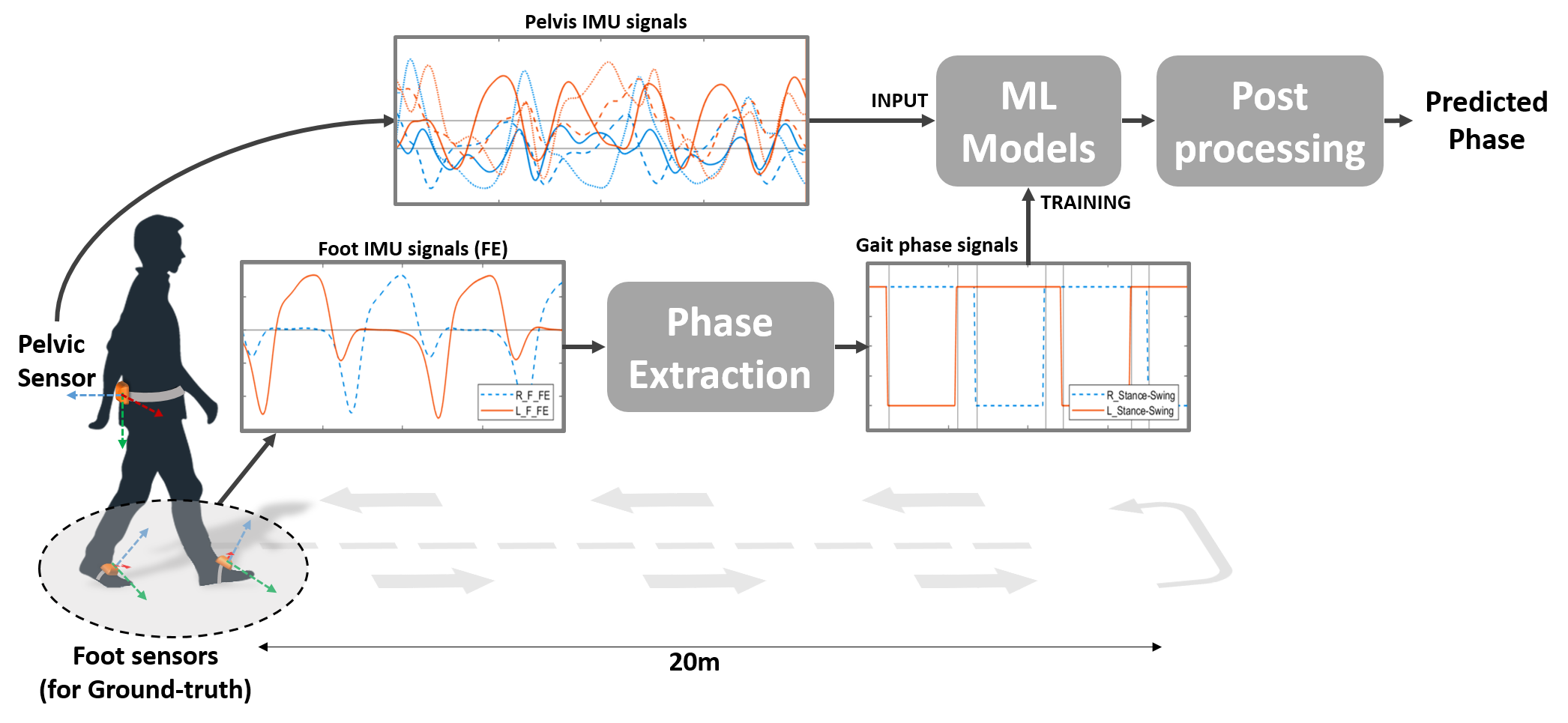}
    \caption{Overview of the proposed phase detection method and the experiment used for data acquisition }
    \label{Experiment}
\end{figure*}

The challenging work of measuring human gait evolved greatly from the traditional visual observation \cite{krebs1985reliability} to the current methods of using threshold, peak detection, handcrafted features, and rule-based methods \cite{kim2017development,oudre2018template,lee2009novel}. The introduction of smaller, lighter, and cheaper sensors like the inertial measurement unit (IMU) sensors made it possible to break free from the laborsome and costly ways of using motion capture systems and force-plates which were limited to a strict clinical setting \cite{barker2006accuracy, winter2009biomechanics}. The advent of machine learning-based methods such as Hidden Markov Models (HMM) \cite{mannini2012gait,taborri2014novel,bae2011gait}, and Support Vector Machines (SVM) \cite{mannini2016machine}, Deep CNN \cite{hannink2016sensor}, and Recurrent Neural Networks (RNN) \cite{lin2021gait} eased the burden of gait measurement further and opened a new horizon of accurate gait assessments. 

When measuring gait, it can be quantified through temporal characteristics in which precise detection of the heel-strike (HS) and toe-off (TO) for each foot matters. From HS and TO detected,  a gait phase gets computed. Although the existing methods have achieved fairly good performance in detecting, the use of multiple sensors and a complex pre-processing process for extracting features have been pointed out as a major hurdle. On top of that, the placement of the sensors mostly on the lower body parts interferes with natural walking and limits its application to the daily life of individuals.

As an alternative to the multiple sensors on the lower body parts, a single sensor on the pelvis can be suggested for its positional advantage and the abundance of information acquirable. The pelvis might be an ideal place for any wearable sensors for it is a common site for wearing a belt and the site hinders common daily activities the least. And the pelvis is a valuable source of information since a single sensor detection of the right and left feet is possible and it is linked to three of the six determinants of gait namely pelvis rotation, pelvic tilt, and lateral displacement of the pelvis \cite{inman1953major}. It is also aligned to the vertical midline of the body at the center of mass (COM) and essentially links the lower limb to the upper body which enables it to transmit force between the two and control whole-body balance. Using the signals from the pelvis, activities can be recognized and even estimating the postures such as sitting and lying is possible. The signals from the pelvis are rich with information about daily activity patterns and carry comprehensive information on gait and motion.  Yet, little attempt has been made in employing the pelvis signals for the accuracy of using them was not comparable to that of using lower body part signals \cite{zijlstra2003assessment, de2019concurrent}.

Thus, this study aimed to explore a way of improving the accuracy of gait event detection in the elderly using a single sensor on the pelvis and deep learning models. The elderly with or without health issues were recruited and their gait was measured. Various deep learning models learned the interrelationship of the gait event information that exists in the pelvis signal and predicted gait events. Then, the prediction was compared with the groundtruth from the sensors on the feet. Suggesting a reliable way of using a single sensor on the pelvis in gait detection, the study is expected to contribute to reducing the burden of gait detection and increase its applicability in future wearable devices that can be used for remote health monitoring (RHM) or diagnosis based thereon. 

\section{Methods}

\begin{table*}[t]\caption{Demographic information of the subjects}
\label{table_Demogr}
\centering
\begin{tabular}{llll}
\toprule
\textbf{Characteristic}           & \textbf{All subjects n=169}   & \textbf{Healthy subjects n=94} & \textbf{Patients n=75}          \\ \midrule
Age (years), Mean $\pm$ SD(Range) & 74.89 $\pm$ 5.08 (60-87)      & 74.66 $\pm$ 4.75 (64-87)       & 75.17 $\pm$ 5.48 (60-87)        \\
Height (cm), Mean $\pm$ SD(Range) & 159.67 $\pm$ 7.23 (141.9-171) & 160.5 $\pm$ 7.01 (141.9-171)   & 155.65 $\pm$ 7.55 (151.3-170.4) \\
Weight (kg), Mean $\pm$ SD(Range) & 61.1 $\pm$ 9.34 (42.3-91)     & 62.01 $\pm$ 9.69 (42.3-91)     & 59.92 $\pm$ 8.79 (42.5-80)      \\ \midrule
Gender                            &                               &                                &                                 \\
 \quad- Male n (\%)                     & 68 (40.24\%)                  & 42 (44.68\%)                   & 26 (34.67\%)                    \\
 \quad- Female n (\%)                   & 101 (59.77\%)                 & 52 (55.32\%)                   & 49 (65.34\%)                    \\ \bottomrule
\end{tabular}
\end{table*}

\subsection{Data-collection}
The subjects of 169 community-dwelling elderly aged between 60 and 80 years were recruited for the study. This study was approved by the Institutional Review Board of Kyung Hee University Medical Center (IRB No. 2017-04-001). Written informed consent was obtained from all the participants before participation in the study. The subjects were divided into healthy and patient groups depending on their health conditions at participating. The patient group included subjects with frailty (n=47), cognitive impairment (n=8), fall history (n=11), and a combination of them(n=9). Frailty was defined as the 5-item FRAIL scale test result of 1 or higher \cite{morley2012simple} and cognitive impairment was defined as the mini-mental state examination (MMSE) \cite{folstein1975mini} scores less than 24. Subjects who answered in the questionnaire that they had a history of falling within the last one year and received hospital treatment for it were included in the patient group. With these criteria, 75 subjects were included in the patient group while the rest of 94 were in the healthy group. All subjects were capable of walking without any help from others or aid from devices at the time of data collection. Table \ref{table_Demogr} summarizes the demographic details of the subjects.

Three commercial IMU sensors were used for this study: one on the pelvis and two on the feet (Xsens MVN, Enschede, and Netherland). An IMU sensor was attached to a belt that the subjects wore around their waist while the other two sensors were tied around the feet one for each foot as depicted in Fig. \ref{Experiment}. Wearing the sensors, the subjects were asked to walk a 10-meter path three times at their preferred speed and faster than their usual speed which made each subject walk the path a total of six times. The three translational and three rotational inertial data were collected at a sampling rate of 100Hz and passed through a 0.5-6Hz band-pass filter to remove high-frequency noise.

\begin{figure}[hbt!]
    \centering
    \captionsetup{justification=justified}
    \includegraphics[width=\linewidth]{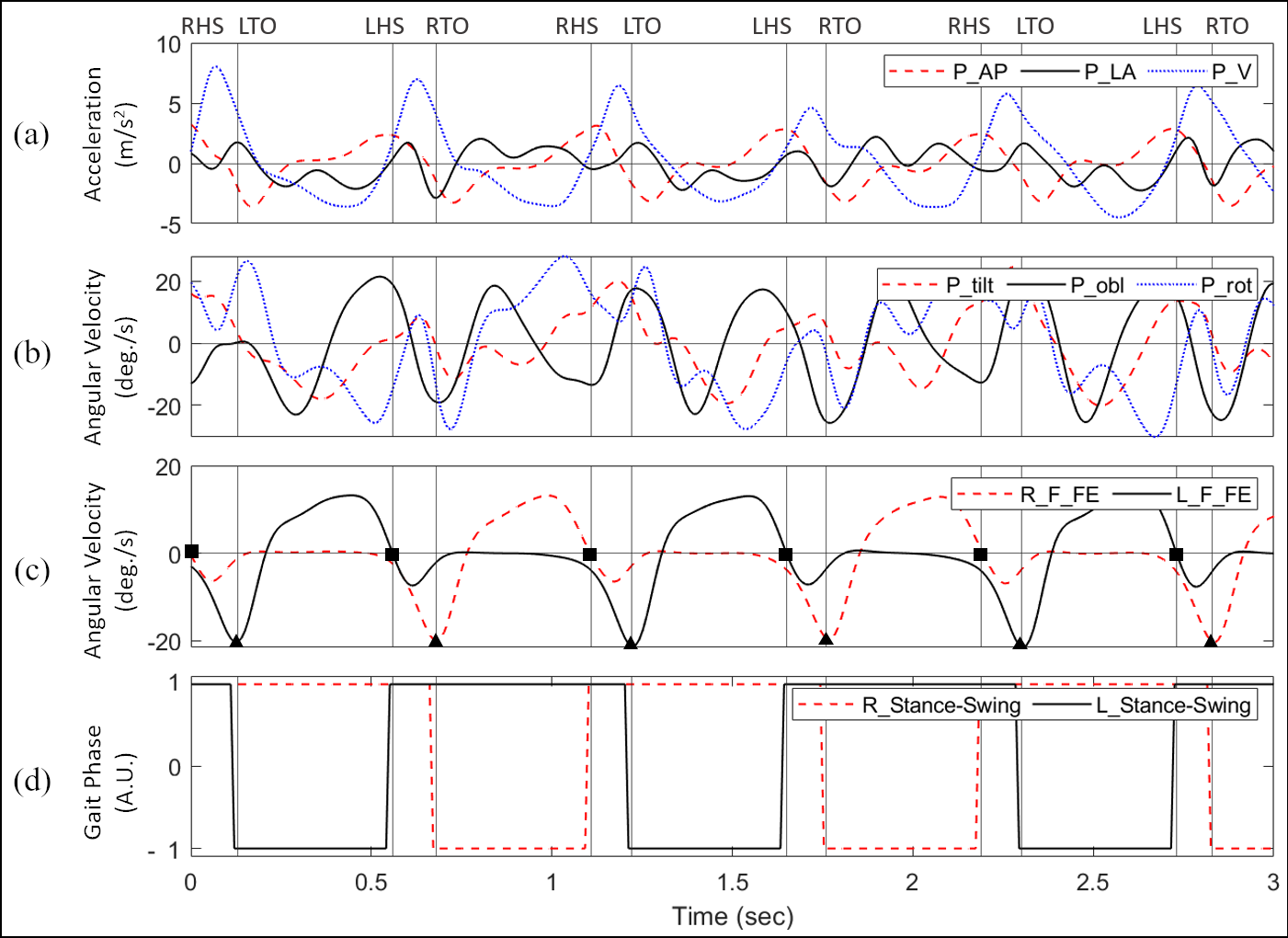}
    \caption{Input signals consist of (a) acceleration signals, (b) angular-velocity signals from the pelvis. (c) Groundtruth is extracted by identifying events from the angular-velocity signals from the feet. (d) The stance and swing phase signals generated for the right and left foot as two continuous groundtruth signals for the regression based models. }
    \label{Signals}
\end{figure}

Figure \ref{Signals}(a) and (b) show the acceleration and angular velocity signals from the pelvis of a healthy subject. To detect the actual HS and TO, the angular velocity signals for flexion and extension of the right and left foot in the sagittal plane were used (Fig. \ref{Signals}(c)). The toe-off events were detected as inverted high-amplitude peaks marked with squares in and the heel-strike events were detected as the zero-crossing before the inverted low-amplitude peaks marked with triangles in the same figure [33].  Using the four events of HS and TO of each foot, the groundtruth data was generated (Fig.\ref{Signals}(d)). The period of the gait cycle with the foot on the ground (HS to TO) was called the stance phase while that with the foot in the air (TO to HS) was called the swing phase [11]. Fig. \ref{Signals}(d) shows the right foot phase signal represented as a dashed red line and the left foot signals as the solid black line.  A value of 1 was assigned for the stance phase and -1 for the swing phase.

\begin{figure*}[hbt!]
  \includegraphics[width=\textwidth]{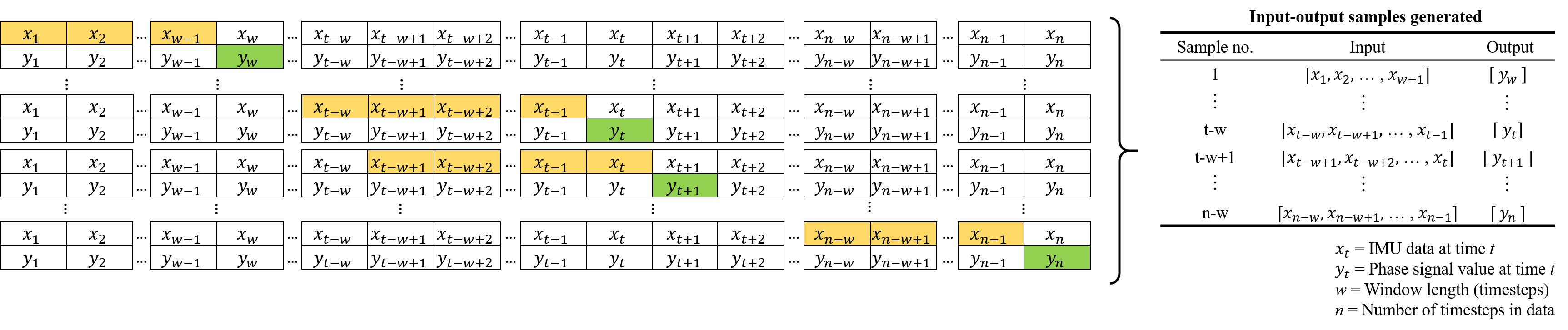}
  \caption{The training data is prepared as input-output pairs where the input consist of previous values of the pelvis IMU signals in a moving window and the output is the next value of groundtruth phase signal after the window.}
  \label{DataPrep}
\end{figure*}

Fig. \ref{DataPrep} illustrates how input-output data pairs were generated for training one-step-ahead prediction where the input $x$ refers to the pelvis IMU data in the sliding window and the output y refers to the right and left phase signal values for the timestep next to the sliding window.  The first pair started with the input of $x_1, x_2, ..., x_{w}$ where $w$ is the window length and with the output of $y_{w+1}$. The window was then shifted by one timestep. Hence for each pair at timestep $t$, the input was $x_{t-w}, x_{t-w+1}, ..., X_{t-1}$ with the output of $y_t$. For the input data of timestep $n$, a total of $n-w$ input-output pairs were generated.

\subsection{deep learning models}
A total of 16 deep learning models were trained and tested. The models included classical, convolutional neural network (CNN), Recurrent Neural Networks (RNN) models with or without the Bidirectional or Attention mechanisms along with some hybrid ones that combined CNN with RNN models with or without the Bidirectional or Attention mechanisms. The classical models included Convolutional Neural Networks (CNN) and Multi-Layer Perceptron (MLP) model while RNN models included Long Short-term Memory (LSTM), Gated Recurrent Unit (GRU), Bidirectional LSTM (BiLSTM), Bidirectional GRU (BiGRU), Stacked LSTM, Stacked GRU, Stacked LSTM with the Attention mechanism (stacked-LSTM-Att), and stacked-GRU with the Attention mechanism (stacked-GRU-Att). The hybrid models included CNN combined with LSTM (CNN-LSTM), CNN combined with GRU (CNN-GRU), CNN combined with Bidirectional LSTM (CNN-BiLSTM), CNN combined with Bidirectional GRU (CNN-BiGRU), CNN combined with Bidirectional LSTM and the Attention mechanism (CNN-BiLSTM-Att), and CNN combined with Bidirectional GRU and the Attention mechanism (CNN-BiGRU-Att).

\begin{figure}[b!]
    \centering
    \captionsetup{justification=justified}
    \includegraphics[width=\linewidth]{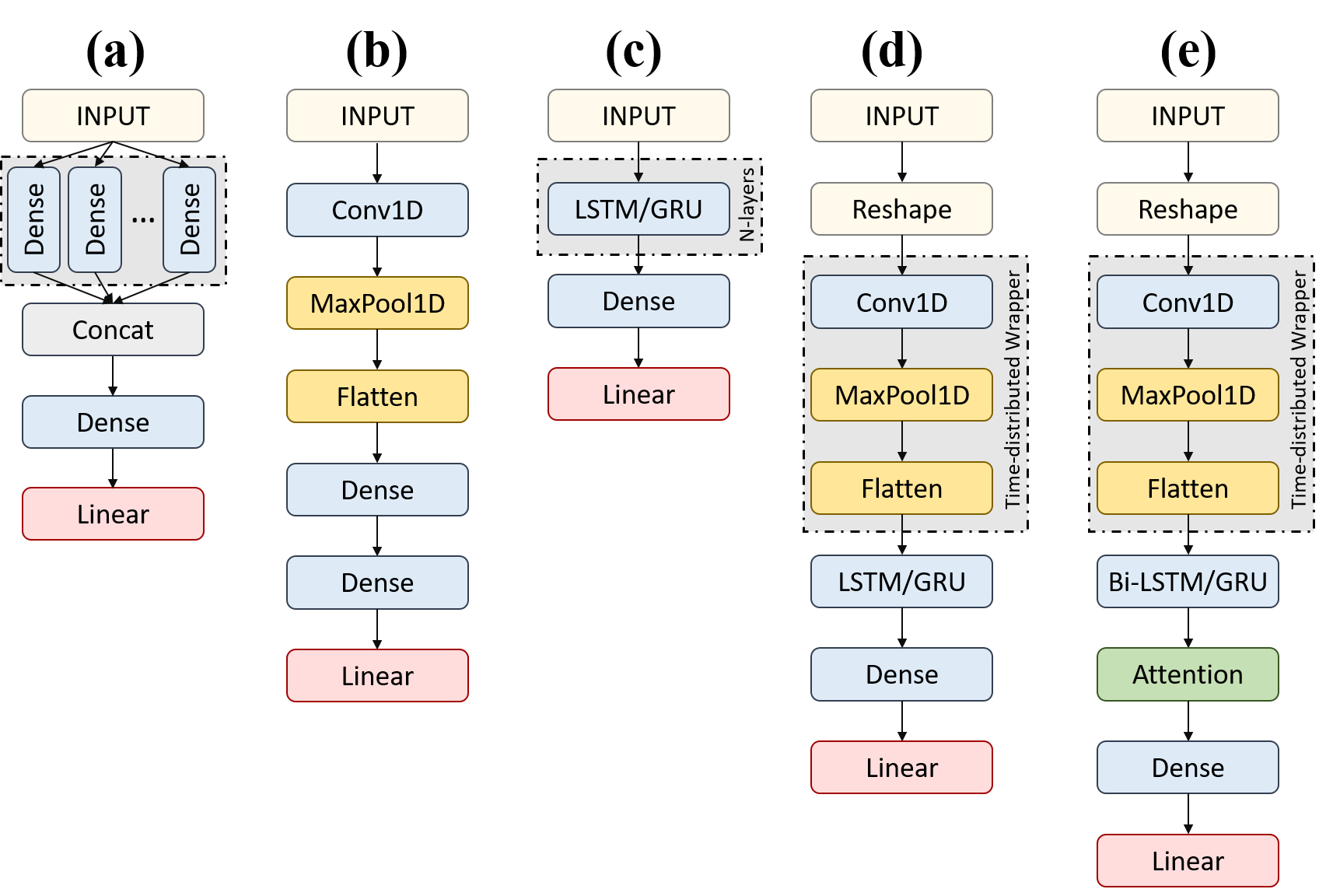}
    \caption{Brief architectural details for models, (a) MLP (b) CNN (c) LSTM and GRU based models (d) CNN-RNN hybrid models (e) CNN-RNN hybrid models with bidirectional and Attention mechanisms}
    \label{NetworksArch}
\end{figure}

For all models, the raw data for training consisted of six IMU signals from the pelvis as inputs and the stance and swing phase signals from both feet as outputs. The models used the generated input and output data samples and made a one-step-ahead prediction. The hyper-parameters for all models were optimized to get the best accuracy for each. A Dense layer followed by a Linear activation function (AF) was used for the final output.

The architecture adopted for the deep learning models is presented in Fig. \ref{NetworksArch}.   Fig. \ref{NetworksArch}(a) shows the architecture for MLP. The model included a Dense layer of 100 neurons for each of the six input signals and a concatenate layer for combining them before connecting to a Dense layer. As for the CNN model, the input was first reshaped to the dimensions [\textit{samples}, \textit{timesteps}, \textit{features}] for making it compatible with the one-dimensional convolutional layer, Conv1D (Fig. \ref{NetworksArch}(b)). The Conv1D layer was followed by a one-dimensional pooling layer, MaxPool1D. After flattening, the outputs went into a Dense layer of 50 neurons and the last Dense layer connected this sequence to the final output after a Linear AF. As for the RNN models with a single or stacked LSTM or GRU with or without attention, the single-layer vanilla LSTM and GRU networks had 100 hidden units and a Dense layer and a Linear AF followed (Fig. \ref{NetworksArch}(C)). The stacked LSTM and GRU included two layers stacked together with 100 hidden units for each. The first layer fed its hidden state to the second one which was used for the output prediction. 

For the hybrid models combining CNN and RNN models, the input with dimensions [samples, window–length, features] was reshaped by splitting the window-length dimension into two segments. With a window-length of 80 timesteps, the input dimensions $[samples, 80, 6]$ were transformed to $[samples, 2, 40, 6]$. The reshaped input was fed to Conv1D. MaxPooling1D followed the convolutional layer, which was then flattened in the end (Fig. \ref{NetworksArch}(d)). The time-distributed layer that wrapped the convolutional blocks enabled applying the same instance of blocks to all the temporal slices of the input \cite{abadi2016tensorflow}. The output from this CNN went into the single layer of LSTM or GRU with 600 hidden units. The Dense layer followed with Linear AF as in other models.  

As for the models with the Bidirectional and Attention mechanisms, the Bidirectional mechanism provided the forwards and backward sequence of the input to the two different RNN layers, allowing the network to make use of past and future context for each point in the input to make predictions \cite{graves2012supervised}. The Self-attention mechanism enabled the model to give more weightage to a specific part of the input sequence \cite{vaswani2017attention}. In this context, the Attention allowed the models to find temporal dependencies in certain periods in the sliding window instead of relying on all to make more accurate predictions. Fig. \ref{NetworksArch}(e) shows CNN-RNN hybrid with the Bidirectional mechanism incorporated into the RNN, and the Attention mechanism that is connected at the output of RNN networks. For the final output, the Attention layer was followed by the Dense layer and Linear AF as others.

All models were trained using Adam optimizer and mean squared error as the loss metric. An early stopping criterion was used to retrieve the best model by minimizing the validation loss with the patience of 10 epochs.

\begin{figure}[hbt!]
    \centering
    \captionsetup{justification=justified}
    \includegraphics[width=\linewidth]{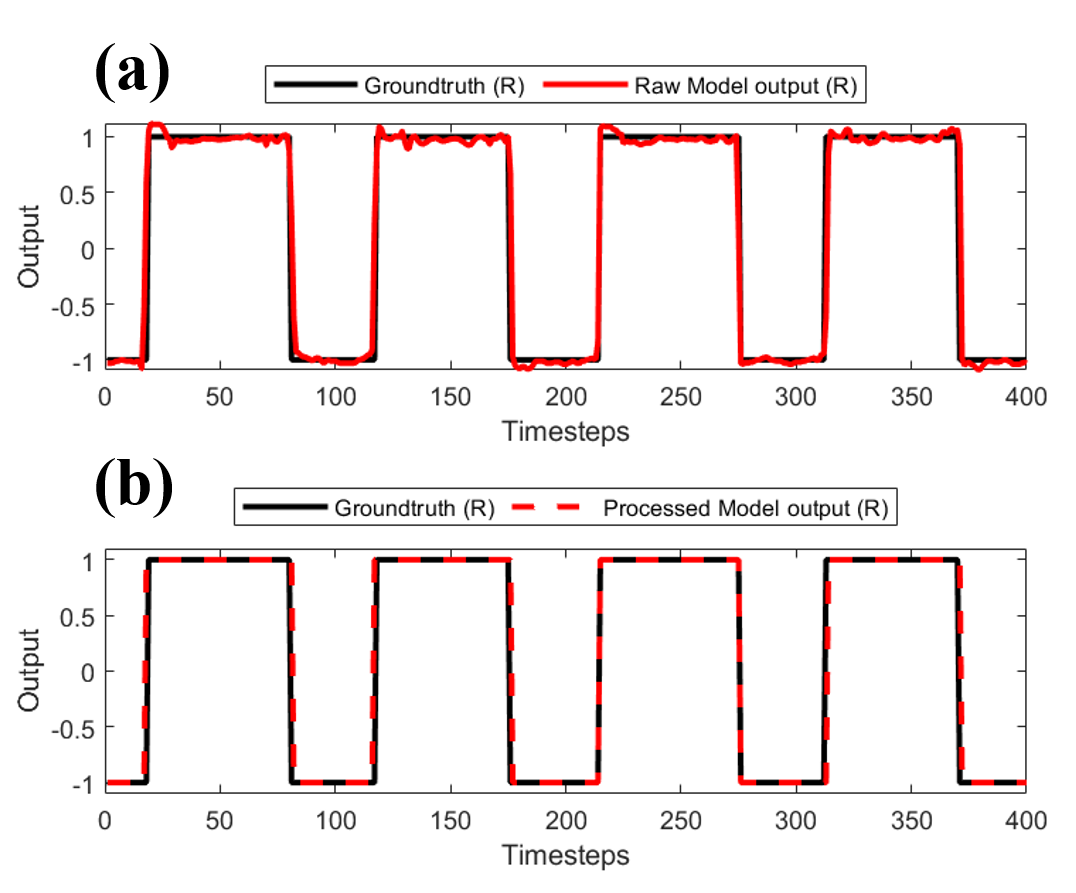}
    \caption{Groundtruth compared to (a) pre-processed output signal and (b) post-processed output signal.}
    \label{OutputVsGd}
\end{figure}

\subsection{Output post-processing}
Post-processing of the output signals was performed to remove the noisy spikes in the raw outputs and improve the prediction accuracy (Fig. \ref{OutputVsGd}). As the first step, all transitions across the zero-line were extracted from the output signals. Then, these transitions were filtered to remove the disturbances which could be mistaken as phase transitions. Since the output signal is essentially a pulse train, the valid phase transitions were identified by distinguishing between noisy spikes and real pulses. For a pulse to be identified as valid, three conditions were used: (1) the maximum value of the pulse must be higher than 0.5, (2) the mean value of the pulse must be higher than 0.6, and (3) the pulse-width must be higher than three timesteps. The processed output matched the groundtruth better (Fig. \ref{OutputVsGd}(b)).

\begin{figure}[hbt!]
    \centering
    \captionsetup{justification=justified}
    \includegraphics[width=\linewidth]{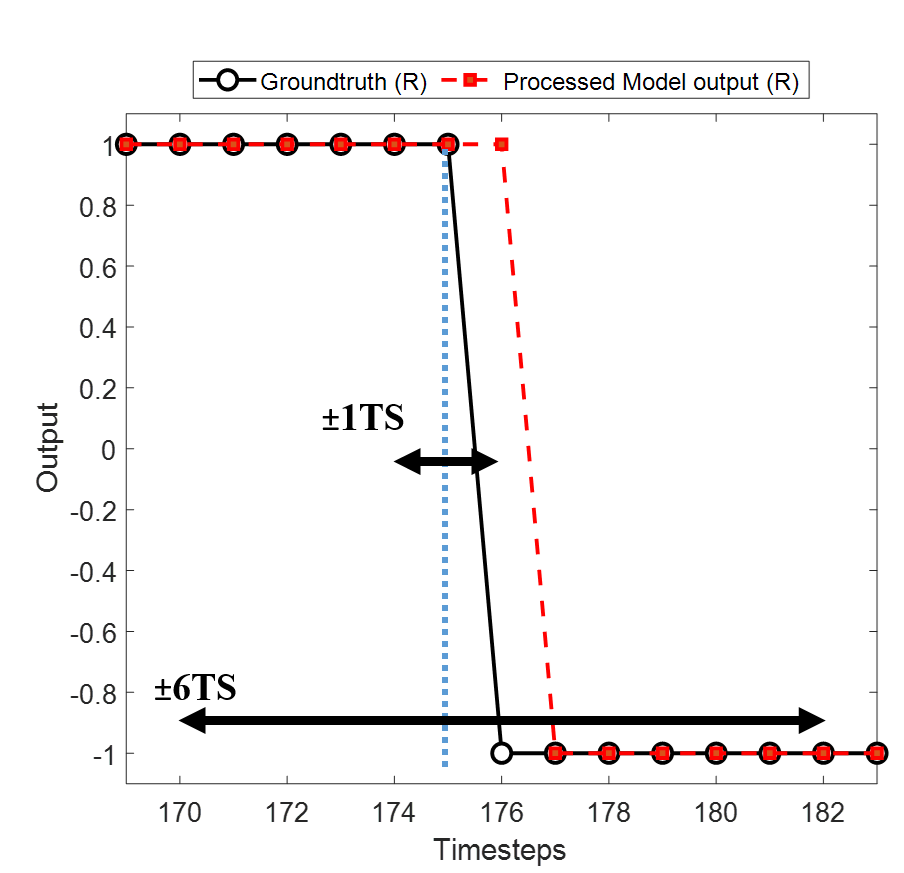}
    \caption{The accuracy is measured with different precision levels termed as tolerance windows $\pm$1TS, $\pm$2TS, ..., $\pm$6TS, where 1TS is a single timestep of 1ms}
    \label{PrecisionLevels}
\end{figure}

\subsection{Accuracy measurement}

The accuracy was measured by different tolerance windows to verify the precision of the detection. An event was defined as successfully detected when the output transition occurred within W, the tolerance window of size (Fig. \ref{PrecisionLevels}). The performance of each model was measured with the tolerance windows $\pm$1TS, $\pm$2TS, ..., $\pm$6TS where 1TS refers to one timestep of 1ms. The overall accuracy was defined as the percentage of correctly detected events to the total number of events. 

To investigate further, the accuracy of the models including the data from the patient group was investigated. The models were trained and tested with the healthy group, trained with the healthy group but tested with the patient group, and trained and tested with both groups.

\begin{table}[]
\caption{Accuracy results for all models}
\resizebox{\columnwidth}{!}{
\begin{tabular}{lllllll}
\toprule
\textbf{Model}      & \textbf{$\pm$1TS} & \textbf{$\pm$2TS} & \textbf{$\pm$3TS} & \textbf{$\pm$TS} & \textbf{$\pm$5TS} & \textbf{$\pm$6TS} \\ \midrule
CNN-BiGRU-Att       & 93.89           & 98.29           & 99.02           & 99.46           & 99.64           & 99.73           \\
CNN-BiLSTM          & 93.68           & 98.28           & 99.14           & 99.48           & 99.68           & 99.76           \\
CNN-BiLSTM-Att      & 93.52           & 98.21           & 99.02           & 99.46           & 99.64           & 99.70           \\
CNN-BiGRU           & 93.27           & 98.19           & 99.10           & 99.41           & 99.56           & 99.73           \\
stacked-LSTM-Att & 92.71           & 97.60           & 98.91           & 99.31           & 99.49           & 99.59           \\
CNN-GRU             & 92.33           & 97.82           & 98.93           & 99.40           & 99.57           & 99.65           \\
CNN-LSTM            & 91.77           & 97.89           & 98.94           & 99.44           & 99.61           & 99.76           \\
stacked-GRU-Att  & 91.67           & 97.54           & 98.68           & 99.30           & 99.59           & 99.65           \\
BiGRU               & 89.99           & 97.08           & 98.76           & 99.35           & 99.55           & 99.64           \\
stacked-GRU      & 88.86           & 96.51           & 98.53           & 99.24           & 99.51           & 99.60           \\
stacked-LSTM     & 88.04           & 96.64           & 98.53           & 99.18           & 99.47           & 99.59           \\
BiLSTM              & 86.54           & 95.98           & 98.17           & 99.02           & 99.35           & 99.51           \\
GRU                 & 78.98           & 92.70           & 96.72           & 98.32           & 99.13           & 99.47           \\
LSTM                & 76.83           & 91.61           & 96.21           & 98.18           & 98.89           & 99.24           \\
MLP                 & 69.55           & 86.57           & 92.88           & 95.73           & 97.05           & 97.53           \\
CNN                 & 68.58           & 87.23           & 94.25           & 96.76           & 98.01           & 98.56  \\ \bottomrule        
\end{tabular}
}
\label{table_ResultsAllModels}
\end{table}

\section{Results}
Table \ref{table_ResultsAllModels} summarizes the average accuracy of the models trained and tested with the healthy group by different sizes of tolerance windows. CNN-BiGRU-Att model using the tolerance window of $\pm$6TS achieved the highest accuracy of 99.73. All hybrid models with the Bidirectional mechanism with or without the Attention mechanism showed accuracy comparable to the accuracy of the best performing CNN-BiGRU-Att.

The accuracy increased as the sizes of the tolerance window increased. With a wider tolerance window, the detection rate increased while the precision decreased. The best accuracy for each model was achieved with the tolerance window of $\pm$6TS. Since all models gave little deviation from the tolerance window greater than $\pm$3TS, the comparisons between the models were made with the accuracy at the tolerance window of $\pm$1TS  which means the most precise detection.

\begin{figure*}[hbt!]
    \centering
    \captionsetup{justification=centering}
    \includegraphics[width=\linewidth]{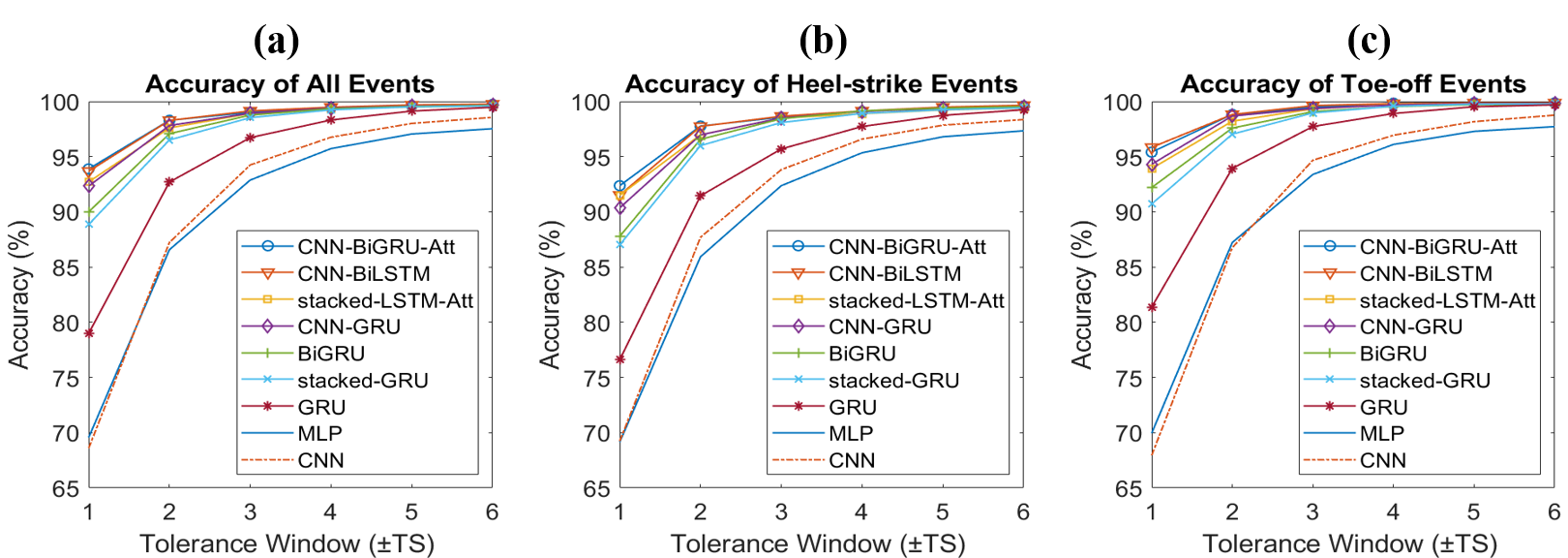}
    \caption{Accuracy plots for key models for (a) all events, (b) HS events, and (c) TO events}
    \label{Results}
\end{figure*}

Fig. \ref{Results} shows the accuracy plots for all events (Fig. \ref{Results}(a)), HS events(Fig. \ref{Results}(b)), and TO events(Fig. \ref{Results}(c)). The nine key models were chosen as the best-performing ones for each type. The events for the right and left foot were averaged for all three plots. In all three events, the accuracy increased as the tolerance window increased. In most models, higher accuracies were achieved with TO events compared to HS events.

\begin{figure}[hbt!]
    \centering
    \captionsetup{justification=centering}
    \includegraphics[width=\linewidth]{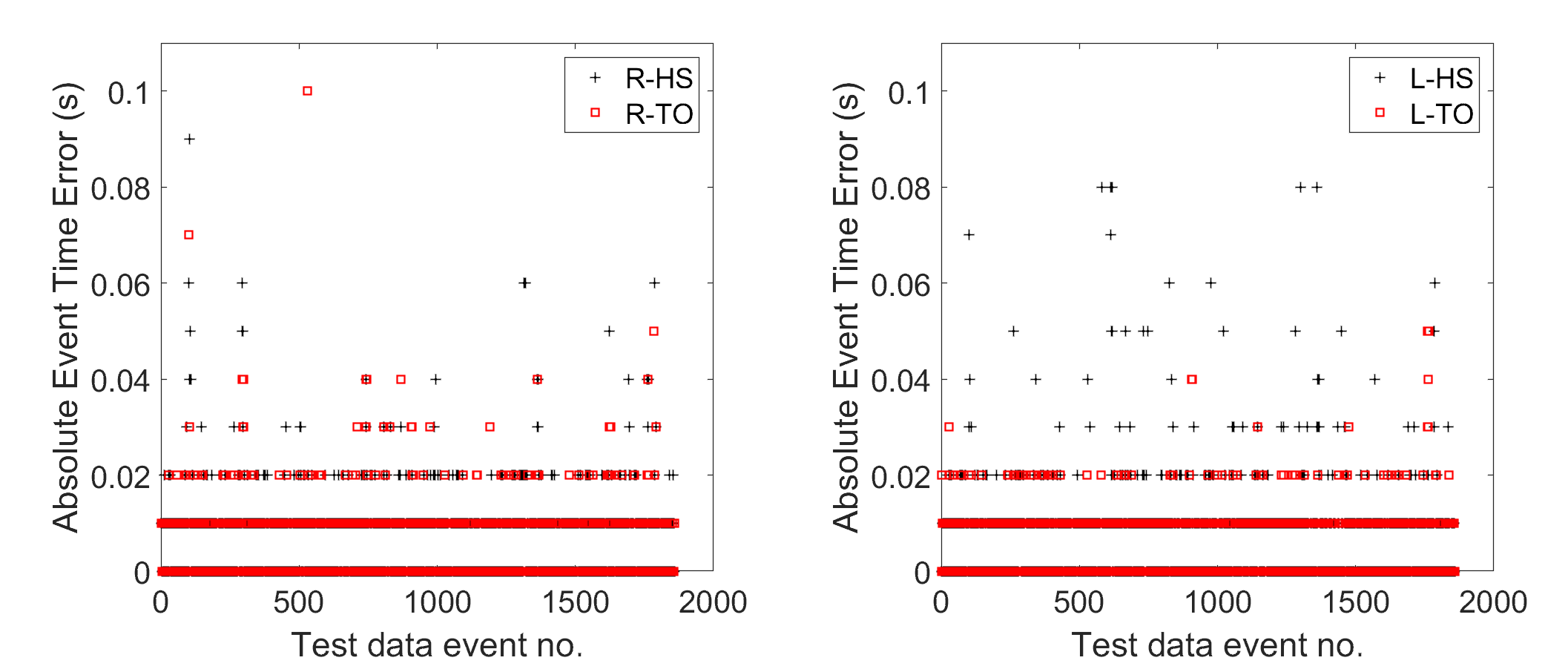}
    \caption{Event prediction errors for right and left, heel-strike and toe-off events}
    \label{PredErrors}
\end{figure} 

% \usepackage{multirow}
\begin{table*}[hbt]
\caption{Results for ablation study with using subset of input signals}
\centering
\begin{tabular}{llllllll}
\toprule
\textbf{No of input signals} & \textbf{Input}                 & \textbf{$\pm$1TS} & \textbf{$\pm$2TS} & \textbf{$\pm$3TS} & \textbf{$\pm$4TS} & \textbf{$\pm$5TS} & \textbf{$\pm$6TS} \\ \midrule
\multirow{6}{*}{1}           & {[}AP{]}                       & 20.57           & 27.75           & 32.12           & 34.65           & 36.26           & 37.52           \\
                             & {[}ML{]}                       & 60.12           & 79.12           & 88.20           & 92.57           & 94.80           & 96.14           \\
                             & {[}V{]}                        & 15.37           & 21.60           & 25.12           & 27.47           & 28.83           & 29.74           \\
                             & {[}TIL{]}                      & 24.52           & 33.52           & 38.65           & 41.89           & 44.03           & 45.42           \\
                             & {[}OBL{]}                      & 15.52           & 24.05           & 31.13           & 36.47           & 40.53           & 43.18           \\
                             & {[}ROT{]}                      & 58.00           & 75.96           & 84.42           & 88.45           & 90.72           & 92.09           \\ \midrule
\multirow{2}{*}{2}           & {[}AP, ML{]}                   & 88.16           & 97.12           & 98.66           & 99.14           & 99.45           & 99.65           \\
                             & {[}AP, ROT{]}                  & 87.23           & 95.45           & 97.61           & 98.66           & 99.27           & 99.45           \\ \midrule
\multirow{2}{*}{3}           & {[}AP, ML, TIL{]}              & 90.15           & 97.28           & 98.68           & 99.21           & 99.60           & 99.70           \\
                             & {[}AP, ML, ROT{]}              & 90.28           & 97.34           & 98.86           & 99.18           & 99.50           & 99.68           \\ \midrule
\multirow{2}{*}{4}           & {[}AP, ML, V, TIL{]}           & 93.40           & 98.25           & 98.95           & 99.30           & 99.60           & 99.68           \\
                             & {[}AP, ML, V, ROT{]}           & 94.11           & 98.48           & 99.13           & 99.46           & 99.68           & 99.79           \\ \midrule
\multirow{3}{*}{5}           & {[}AP, ML, V, TIL, OBL{]}      & 87.97           & 96.57           & 98.31           & 99.14           & 99.49           & 99.62           \\
                             & {[}AP, ML, V, TIL, ROT{]}      & 93.89           & 98.04           & 98.97           & 99.53           & 99.73           & 99.84           \\
                             & {[}AP, ML, V, OBL, ROT{]}      & 93.99           & 98.28           & 98.94           & 99.38           & 99.68           & 99.81           \\ \midrule
6                            & {[}AP, ML, V, TIL, OBL, ROT{]} & 93.89           & 98.29           & 99.02           & 99.46           & 99.64           & 99.73           \\ \bottomrule
\end{tabular}
\label{table:ResultsLimitedSignals}
\end{table*}

To observe the differences in the performance of the events from the right and left foot, the absolute errors in timesteps between the predicted and the groundtruth events were computed. Fig. \ref{PredErrors} shows the event prediction error of CNN-BiGRU-Att. The mean absolute error (MAE) for all events was 5.77ms whereby those for HS and TO events were 6.239ms and 5.24ms, respectively. The TO events showed fewer errors than the HS events. No significant difference was found between using the right foot events and left foot events. 

% \usepackage{multirow}
\begin{table*}[t]
\caption{Accuracy results when patient data is used}
\centering
%\resizebox{\columnwidth}{!}{
\begin{tabular}{lllllllll}
\toprule
\textbf{Input signals}                                              & \textbf{Training} & \textbf{Testing} & \textbf{$\pm$1TS} & \textbf{$\pm$2TS} & \textbf{$\pm$3TS} & \textbf{$\pm$4TS} & \textbf{$\pm$5TS} & \textbf{$\pm$6TS} \\ \midrule
\multicolumn{1}{c}{\multirow{3}{*}{{[}AP, ML, V, TIL, OBL, ROT{]}}} & HS                & HS               & 93.89           & 98.29           & 99.02           & 99.46           & 99.64           & 99.73           \\
\multicolumn{1}{c}{}                                                & HS                & P                & 63.10           & 84.09           & 93.30           & 96.94           & 98.44           & 99.05           \\
\multicolumn{1}{c}{}                                                & Mixed             & Mixed            & 93.63           & 98.04           & 98.85           & 99.21           & 99.44           & 99.59           \\ \midrule
\multirow{3}{*}{{[}AP, ML, V, ROT{]}}                               & HS                & HS               & 94.11           & 98.48           & 99.13           & 99.46           & 99.68           & 99.79           \\
                                                                    & HS                & P                & 62.78           & 83.63           & 92.40           & 96.25           & 97.97           & 98.69           \\
                                                                    & Mixed             & Mixed            & 92.80           & 97.91           & 98.95           & 99.30           & 99.52           & 99.66       \\ \bottomrule   
\end{tabular}
%}
\label{table:ResultsTestingOnPatients}
\end{table*}
An ablation study with subsets of the 6 input signals was performed to investigate the prediction accuracy by the number of input signals. Table \ref{table:ResultsLimitedSignals} shows the list of the best-performing input combinations for the CNN-BiGRU-Att model. Using either ML or ROT and using both of them increased the accuracy. When two input signals were used, the combination of AP and ML and that of AP and ROT gave the accuracy of about 88\% at the tolerance window of $\pm$1TS. The highest accuracy of over 94\% was achieved by using four input signals of AP, ML, V, and ROT. When all six input signals were used, an accuracy of 93.89\% was achieved.

\begin{figure*}[hbt!]
    \centering
    \captionsetup{justification=justified}
    \includegraphics[width=\linewidth]{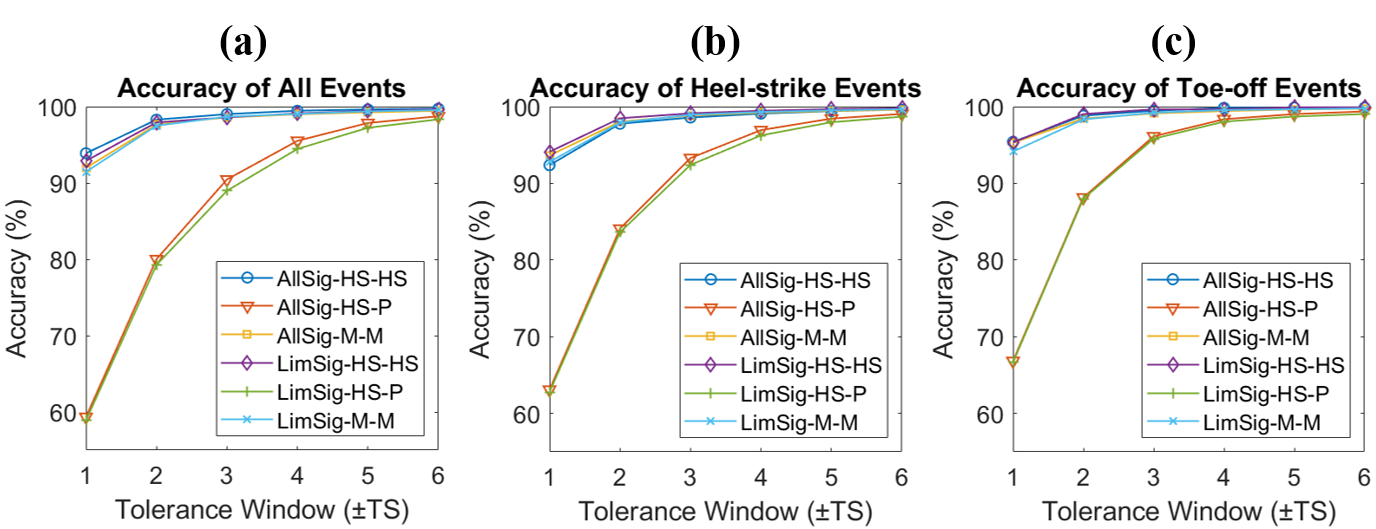}
    \caption{Accuracy plots for (a) all events, (b) heel-strike events, and (c) toe-off events when working with patient data/ The plots includes results from models trained with all 6 pelvis signals (AllSig) and with limited 4 signals (LimSig). The training and testing includes healthy-trained, patient-tested (HS-P), healthy-trained, healthy-tested (HS-HS), and mixed-trained, mixed-tested (M-M).}
    \label{ResultsPatients}
\end{figure*}

Table \ref{table:ResultsTestingOnPatients} shows the accuracy of the CNN-BiGRU-Att model trained and tested with the same or different subject groups. When all six input signals were used, the models trained and tested with the healthy group exhibited an accuracy of 93.89\% at the tolerance window of $\pm$1TS. When trained with the healthy group but tested with the patient group, an accuracy of 63.10\% was achieved. When trained and tested with both groups, the models achieved an accuracy of 93.63\%. When four input signals of AP, ML, V, and ROT were used, the models trained and tested with the healthy group achieved the highest accuracy of 94.11\% at the tolerance window of $\pm$1TS which was higher than that of using all six signals. When the model was trained with both groups, using all six inputs achieved higher accuracy at the tolerance window of $\pm$1TS and $\pm$2TS. To observe the accuracies of the HS and TO events for these results, accuracy plots are given in Fig. \ref{ResultsPatients}. As observed earlier, the TO events were identified with a higher accuracy than the HS events. 

\section{Discussion}
This study aimed to explore a way of improving the accuracy of gait event detection among the elderly using a single sensor on the pelvis and deep learning technology. A total of 16 models were trained and tested to predict the gait events of the elderly and their prediction was compared with the groundtruth acquired from the feet. The CNN-BiGRU-Att model achieved the highest accuracy of 99.73\% at the tolerance window of $\pm$6TS which was an accuracy comparable to that of using multiple sensors on the lower body parts. An ablation study showed using four signal sets of AP, ML, V, and ROT achieved an accuracy of over 94\% at the tolerance window of $\pm$1TS. The study pioneered the utilization of deep learning technology in predicting gait events using the data from the pelvis and suggested a reliable way of using a single sensor on the pelvis in gait event detection among the elderly. The findings are expected to contribute to reducing the burden of gait measurement and increase the potential of various future technologies being incorporated with the suggested method.

The use of  CNN  together with RNN models improved the prediction accuracy since CNN first extracted effective features and then the following RNN models could process these features sequentially. The added Bidirectional mechanism took into account both forward and backward temporal perspectives which improved the accuracy further. Adding the Attention mechanisms to the CNN-BiGRU increased the accuracy even more since the Attention mechanisms could focus on the timesteps that were more relevant for the output prediction. Being unable to utilize the temporal nature of the information in the sliding window, MLP and CNN were not able to predict the phase transitions in the output signal accurately until $\pm$4TS although their accuracy improved as the tolerance window sizes grew bigger. 

Many previous studies have looked at the use of multiple sensors to analyze gait events. Lin et al. \cite{lin2021gait} used LSTM-based regression model using five IMUs, two on the thighs, two on the shins, and one on the left shoe. They achieved the mean error (ME) of 2ms for HS, however large errors for TO were reported with the ME of -18ms. In \cite{hannink2016sensor}, Hannink et al. used a deep CNN-based network with input from two inertial sensors placed on the feet and predicted different gait parameters including the HS and TO events as output for which they reported with errors of ±70ms and ±120ms respectively. Sarshar et al. \cite{sarshar2021gait} who also used RNN to train two IMU sensors attached to the shanks, reported an accuracy of 0.9977 for both HS and TO events, however, they did not compute the error in prediction delays. Utilizing rule-based algorithms, Fadillioglu et al. \cite{fadillioglu2020automated} presented an automated gait event detection method using a gyroscope attached to the right shank and reported an MAE of 7ms and 19ms for the HS and TO events respectively. More recently, Yu et al \cite{yu2021gait} also used a single sensor on the right foot with an LSTM-HMM hybrid model for gait event detection, reporting the accuracy only without mentioning the delays. They reported accuracy of 0.9679 and 0.9846 for the HS and TO events. But since both of these works used only a single sensor on the right side, they were unable to get events for the left side. In comparison, only a single sensor on a comparatively less accurate position of the waist was used in the proposed method but much superior overall performance was shown. The studies using multiple sensors not only hinder the natural gait but are also far from practical usage of these methods in daily living. If a single sensor is attached to one side of the limb, the extraction of gait events from the other side is not possible, hence for a complete gait event information two sensors are required for any location on the lower limb except for the waist.  Even though waist is a less accurate position as compared to the lower limb, the proposed method in the current study has still managed to achieve more accurate gait event detection using a single sensor.

Compared with previous attempts that detected gait events from a single sensor on the waist, the proposed CNN-BiGRU-Att model achieved far advanced accuracy. According to a recent survey of gait event detection methods using an IMU sensor mounted on the waist, Gonzalez et al. \cite{gonzalez2010real} used a rule-based method to achieve the lowest MAE of 15ms and 9ms for the HS and TO events respectively. McCamley et al. \cite{mccamley2012enhanced} proposed a Gaussian CWT-based gait event estimation method using a single inertial sensor on the waist, reporting an MAE of 19ms and 32ms for HS and TO respectively. Soaz et al. \cite{soaz2015step} also used a rule-based method with a single waist sensor to assess the gait of elderly for their experiment and reported an error of 20ms for HS. Apart from relatively higher errors than the proposed method, these studies also did not have enough subjects to show the generalizability of their method.

\begin{figure*}[!t]
    \centering
    \captionsetup{justification=centering}
    \includegraphics[width=\linewidth]{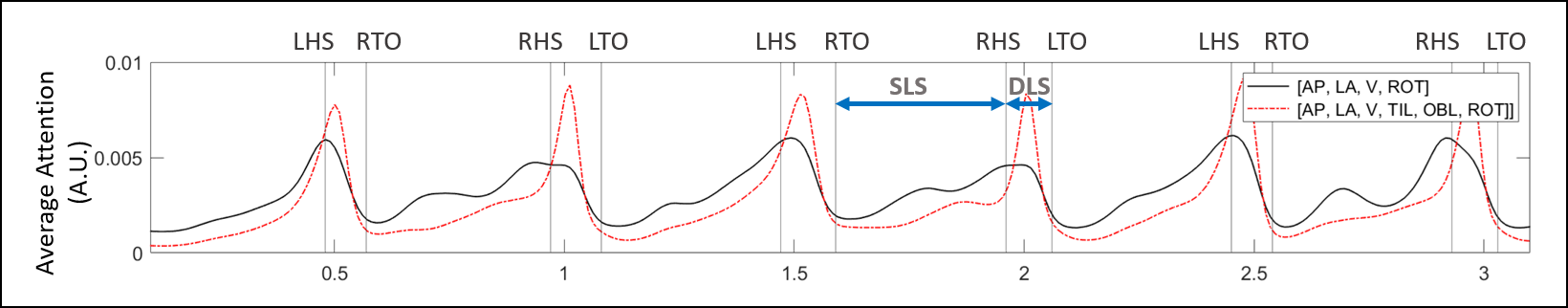}
    \caption{Average Attention plotted for the stacked-LSTM model using all 6 input pelvis signals and limited 4 input signals. The models put more attention to DLS phase between right HS and left TO as compared to the SLS between right TO and right HS.}
    \label{Attention}
\end{figure*} 

The study found that the TO events show fewer errors than the HS events.  The occurrence of extrema in the pelvis signals around these events can serve as a possible explanation. As presented in Fig. \ref{Signals}, the TO events were more aligned with the peaks in pelvic signals such as V, TIL, and ROT which were followed by the  TO events. These characteristics may have contributed to the better performance of the TO events compared with the HS events. 

As for the single input signals used, ML and ROT showed outstanding performance since they are rich in information about right and left. The signals AP and V are the clearest but void of right and left information, so when the prediction was made based on only either of them, they exhibited the lowest performance.  Combining AP with either ML or ROT might have incorporated the movement information on both forward and backward with that on right and left and resulted in fairly improved accuracy. The overall accuracy of using AP, ML, V, and ROT was higher than the accuracy of using all six signals, probably because of the bigger variation found in TILT and OBL among the elderly \cite{perry2010gait}. AP, ML, and V being acceleration signals may have contributed as well since the accelerometer output is generally less prone to sensor location errors compared to a gyroscope. 

When the model was trained and tested with the same and different groups, the models trained with the healthy group but tested with the patient group exhibited an accuracy of 63.10 at $\pm$1TS with all the six input signals used. Trained with the healthy group, the model was not familiar with the variations and abnormalities that the patient group had but its accuracy improved as the tolerance window sizes grew bigger. The inclusion of TILT and OBL in the healthy group did not significantly improve the accuracy of the model, whereas the inclusion of them in the patient group improved the accuracy, probably because of the variation in all signals being greater in the patient group. It would be advisable to consider using all six signals for patients whose signals show a large variation even for the signals considered relatively clear. 

To investigate how the Attention weight varied when there were all six input signals and when they were limited to the four,  the value of average attention weight for each timestep was plotted using the stacked-LSTM-Att model (Fig. \ref{Attention}).  When the model used all six input signals, it gave much higher attention to the double-limb-support (DLS) phase between the HS and TO events than to the single-limb-support (SLS) phase. When four signals were used, slightly more attention was paid to the SLS phase with its peak attention in the DLS phase lower than that of using all six signals. 

One of the limitations of this study is that the proposed method could not be evaluated on other physical impairments like osteoarthritis, skeletal deformities and neurological diseases such as hemiplegia, Parkinson’s disease, Huntington’s disease, and Alzheimer’s disease. Furthermore, the gait data has not been acquired through long-term or continuous monitoring of subjects in their natural environment and everyday lives. Therefore, our future work will include gait event detection for real-world walking in non-conventional environments and under unconstrained and uncontrolled conditions. 
% Having accurate gait events from a single sensor will open new possibilities for remote health monitoring, activities of daily living (ADL), rehabilitation, and gait assessments out of the clinical environment.

\section{Conclusion}
The study proposed deep learning-based gait detection as a novel and reliable way of using a single sensor on the pelvis in detecting the gait of the elderly. A total of 16 models including CNN, RNN, and CNN-RNN hybrid with or without the Bidirectional and Attention mechanism were trained and tested and a fairly high accuracy of 99.73\% and 93.89\% was achieved by the CNN-BiGRU-Att model at the tolerance window of $\pm$6TS and $\pm$1TS respectively. Advancing from the previous studies exploring gait event detection, the model showed a great improvement in terms of its prediction error having an MAE of 6.239ms and 5.24ms for HS and TO events respectively. For healthy subjects, using all three acceleration signals with ROT as input signals exhibited better performance compared to using all six signals while the performance of using all six signals was better for the patients. Suggesting that reliable gait detection is possible from a single sensor on the pelvis, the study is expected to contribute to lowering the burden of gait detection and expand the applicability of gait detection in future wearable devices.

\section*{Acknowledgment}

This research was supported by the Korea Institute of Science and Technology Institutional Program (Project No. 2E31051) and the Korea Medical Device Development Fund grant funded by the Korea government (the Ministry of Science and ICT, the Ministry of Trade, Industry and Energy, the Ministry of Health \& Welfare, the Ministry of Food and Drug Safety) (Project No. 1711139131).

\printbibliography
\end{document}